# Linear Model Estimators and Consistency under an Infill Asymptotic Domain

Cory W. Natoli[1], Edward D. White[1*], Beau A. Nunnally[1], Alex J. Gutman[2], and Raymond R. Hill[2]

1 Department of Mathematics and Statistics, Air Force Institute of Technology, Wright-Patterson Air Force Base, OH

2 Department of Operational Sciences, Air Force Institute of Technology, Wright-Patterson Air Force Base, OH

* Corresponding author.

**Abstract**

Functional data present as functions or curves possessing a spatial or temporal component. These components by nature have a fixed observational domain. Consequently, any asymptotic investigation requires modelling the increased correlation among observations as density increases due to this fixed domain constraint. One such appropriate stochastic process is the Ornstein-Uhlenbeck process. Utilizing this spatial autoregressive process, we demonstrate that parameter estimators for a simple linear regression model display inconsistency in an infill asymptotic domain. Such results are contrary to those expected under the customary increasing domain asymptotics. Although none of these estimator variances approach zero, they do display a pattern of diminishing return regarding decreasing estimator variance as sample size increases. This may prove invaluable to a practitioner as this indicates perhaps an optimal sample size to cease data collection. This in turn reduces time and data collection cost because little information is gained in sampling beyond a certain sample size.

# Linear Model Estimators and Consistency under an Infill Asymptotic Domain

## 1. INTRODUCTION

Methods for representing data by functions or curves have received much attention in recent years (Górecki et al. 2018). Such data are known in the literature as functional data (Ramsay and Silverman 2005; Ferraty and Vieu 2006; Horváth and Kokoszka 2012). Functional data can have a spatial component because data are collected somewhere at some location and time (Haining 2003). Due to correlation among observations, the modelling of covariance is one of the most studied subjects in spatial statistics (Cressie 1993; Stein 1999; Genton and Kleiber 2015). For finite-dimensional data, the exponential, the Gaussian, and the Matérn are examples of parametric covariance function models often adopted.

These covariance functions can model correlated observations since such functions account for the spatial distance between points, however, they generally still assume an increasing time domain to establish consistency for parameter estimators. Because functional data generally possess a finite-dimensional observational window, an increasing time domain is not always appropriate to adopt since increasing sample size entails adding more points between already existing observations. Consequently, a fixed time domain, referred to either as fixed-domain asymptotics (Stein 1999) or infill asymptotics (IA) (Cressie 1993), is more appropriate. Unlike the customary consistency of parameter estimators under an increasing time domain, an IA domain is unlikely to reach similar consistency conclusions (Morris and Ebey 1984; Lahiri 1996; White 2001;

Mills 2010).

Investigating estimator consistency under an IA domain is the goal of this paper. We explore the effect that the IA domain has on the properties of simple linear regression estimators. Specifically, the variance of the estimators for the intercept and slope parameters of a line. Section 2 provides a brief background on IA, the adopted Ornstein-Uhlenbeck process, and the framework of the linear model used. In Sections 3-5, we derive the true variance of the maximum likelihood estimators, the derivatives of the variances, the derivatives' asymptotic properties, and the inconsistency for each of these parameter estimators. In Section 6, we discuss our overall findings and how these results can be extended to any smooth continuous function that can be approximated by piecewise linear functions (Sontag 1981).

## 2. BACKGROUND

Infill asymptotics presents a unique challenge in that the finite domain (in either time or space) results in an increased dependence among samples as more samples are gathered. Under this fixed domain, samples are constrained by a boundary and therefore the density of observations increases proportionately to the increasing number of observations sampled. This effect results in the observations becoming increasingly closer together (i.e., denser), which in turn increases stochastic dependency. Mills (2010) remarks that the maximum distance between two observations is bounded and defines infill asymptotics as:

Definition – *Suppose sampling within the domain of a process were to occur in a manner which spreads the samples as far apart as possible. If:*

$$\lim_{n \to \infty} \max_{i,j \, \in \{1,2,\dots,n\}} |t_i - t_j| = C$$

where $C \in \mathcal{R}^+$ is a finite constant, $i, j \in \mathbb{Z}^+$ are indices which order the samples, and $t_i, t_j \in \mathcal{R}^+ \cup \{0\}$ are time points corresponding to the $i^{th}$ and $j^{th}$ indices, respectively, then the domain of the process is an Infill Asymptotics (IA) domain.

Recognizing this fixed observational domain is critical in properly accounting for the dependence among samples. One common method for modelling the covariance among IA samples is via the Ornstein-Uhlenbeck (OU) process (Parzen 1962). An OU process is a Gaussian process satisfying: $E(\varepsilon_{t_i}) = 0$ and $Cov\left(\varepsilon_{t_i}, \varepsilon_{t_j}\right) = \sigma^2 e^{-\lambda|t_i - t_j|}$, where $\sigma^2 > 0$ is the overall variance, $\lambda > 0$ is a tuning parameter, and $\varepsilon_{t_i}, \varepsilon_{t_j}$ are two random observations in time. For this article, we treat $\lambda$ as an unknown, fixed variable and provide results as a function of $\lambda$. An OU process is also known as a spatial autoregressive process (Cressie 1993), an exponential covariance function (Stein 1999), or a special case of the Matérn covariance function (Mills 2010).

For this article, suppose that an evenly spaced sampling method is used for collecting observations over a fixed interval [0, T] with the first observation at $t_1 = 0$ and the final observation taken at $t_n = T$. Without loss of generality, it can be assumed the observations are sampled uniformly over the fixed interval [0,1] and $\sigma^2$ is known and equal to 1. Using these assumptions, we have $E(\varepsilon_{t_i}) = 0$ and $Cov\left(\varepsilon_{t_i}, \varepsilon_{t_j}\right) = e^{-\lambda|i-j|/(n-1)}$, with $\lambda > 0$ and $i, j \in [1, n]$. This can be rewritten as $Cov\left(\varepsilon_{t_i}, \varepsilon_{t_j}\right) = \rho^{|i-j|}$ where $\rho = e^{-\lambda/(n-1)}$. Therefore, the variance-covariance matrix of $n$ observations, $\mathbf{\Sigma}$, and its inverse, $\mathbf{\Sigma}^{-1}$, are defined throughout this article as follows:

$$\Sigma = \begin{bmatrix} 1 & \rho & \rho^2 & \cdots & \cdots & \rho^{n-1} \\ \rho & 1 & \rho & \cdots & \cdots & \vdots \\ \rho^2 & \rho & \ddots & \ddots & \ddots & \vdots \\ \vdots & \vdots & \ddots & \ddots & \rho & 0 \\ \vdots & \vdots & 0 & \rho & 1 & \rho \\ \rho^{n-1} & \cdots & \cdots & \rho^2 & \rho & 1 \end{bmatrix}$$

$$\Sigma^{-1} = \frac{1}{1-\rho^2} \begin{bmatrix} 1 & -\rho & 0 & \cdots & 0 & 0 \\ -\rho & 1+\rho^2 & -\rho & 0 & \vdots & 0 \\ 0 & -\rho & 1+\rho^2 & \ddots & 0 & \vdots \\ \vdots & 0 & \ddots & \ddots & -\rho & 0 \\ 0 & \vdots & 0 & -\rho & 1+\rho^2 & -\rho \\ 0 & 0 & \cdots & 0 & -\rho & 1 \end{bmatrix}$$

For model specification, we adopt the general linear model as customarily introduced in a beginning regression course. Specifically, we adopt (1) to establish the framework of parameter estimation with the assumption $E[\boldsymbol{\varepsilon}] = 0$ and $V[\boldsymbol{\varepsilon}] = \boldsymbol{\Sigma}$.

$$\boldsymbol{Y} = \boldsymbol{X}\boldsymbol{\beta} + \boldsymbol{\varepsilon} \qquad (1)$$

To determine the maximum likelihood estimator (MLE) of $\boldsymbol{\beta}$ under the scenario where $\boldsymbol{\Sigma}$ is a known, positive definite variance-covariance matrix, we assume $\boldsymbol{Y}$ is distributed as a multivariate normal. Consequently, the likelihood is:

$$L(\boldsymbol{\beta}|\boldsymbol{Y},\boldsymbol{\Sigma}) = |2\pi\boldsymbol{\Sigma}|^{-\frac{1}{2}} \exp\left[\left(-\frac{1}{2}\right)(\boldsymbol{Y} - \boldsymbol{X}\boldsymbol{\beta})'\boldsymbol{\Sigma}^{-1}(\boldsymbol{Y} - \boldsymbol{X}\boldsymbol{\beta})\right]$$

Maximizing the likelihood function with respect to $\boldsymbol{\beta}$ is equivalent to maximizing the loglikelihood due to the invariance principle of maximum likelihood estimators.

$$l(\boldsymbol{\beta}|\boldsymbol{Y},\boldsymbol{\Sigma}) \propto -\log|\boldsymbol{\Sigma}| - (\boldsymbol{Y} - \boldsymbol{X}\boldsymbol{\beta})'\boldsymbol{\Sigma}^{-1}(\boldsymbol{Y} - \boldsymbol{X}\boldsymbol{\beta})$$

Maximizing the loglikelihood is equivalent to minimizing:

$$Q = (\boldsymbol{Y} - \boldsymbol{X}\boldsymbol{\beta})'\boldsymbol{\Sigma}^{-1}(\boldsymbol{Y} - \boldsymbol{X}\boldsymbol{\beta}) = \boldsymbol{Y}'\boldsymbol{\Sigma}^{-1}\boldsymbol{Y} - 2\boldsymbol{\beta}\boldsymbol{X}\boldsymbol{\Sigma}^{-1}\boldsymbol{Y} + \boldsymbol{\beta}\boldsymbol{X}'\boldsymbol{\Sigma}^{-1}\boldsymbol{X}\boldsymbol{\beta}$$

The minimum is found by taking the respective derivative of $Q$, setting it to 0, and solving. The second derivative is also taken to ensure a minimum is found. The derivative is: $\frac{dQ}{d\boldsymbol{\beta}} = -2X'\Sigma^{-1}Y + 2X'\Sigma^{-1}X\widehat{\boldsymbol{\beta}}$ and the second derivative is: $\frac{d^2Q}{d\boldsymbol{\beta}^2} = 2X'\Sigma^{-1}X \geq \mathbf{0}$. Thus, the MLE is defined as:

$$\widehat{\boldsymbol{\beta}} = \frac{X'\Sigma^{-1}Y}{X'\Sigma^{-1}X} \qquad (2)$$

This estimator is unbiased since:

$$E[\widehat{\boldsymbol{\beta}}] = E[(X'\Sigma^{-1}X)^{-1}X'\Sigma^{-1}Y] = (X'\Sigma^{-1}X)^{-1}X'\Sigma^{-1}E[Y] = (X'\Sigma^{-1}X)^{-1}X'\Sigma^{-1} = \boldsymbol{\beta}$$

with variance:

$$Var(\widehat{\boldsymbol{\beta}}) = Var[(X'X)^{-1}X'Y] = (X'X)^{-1}X'Var(Y)((X'X)^{-1}X')'$$

$$= (X'\Sigma^{-1}X)^{-1} \qquad (3)$$

The estimators calculated here and used in the article are the best linear unbiased estimator (BLUE) for $\boldsymbol{\beta}$ via the Gauss Markov theorem (Shaffer 1991) and are the uniformly minimum-variance unbiased estimators (UMVUE) (Berger 2002).

## 3. INTERCEPT ONLY MODEL

The intercept model is the simplest form of (1) and provides a foundation from which to extend to more complex models. Some of the results presented in this section mirror that of White (2001), specifically the derived form of the variance estimator and its limiting value. For this section and subsequent ones, we present results for when $n > 2$ since $n$ must be greater than 1 for variability to exist and $n = 2$ results in a straight-forward form.

Starting with (1), let

$$Y = \begin{bmatrix} Y_{t_1} \\ Y_{t_2} \\ \vdots \\ \vdots \\ Y_{t_{n-1}} \\ Y_{t_n} \end{bmatrix}_{nx1} \quad X = \begin{bmatrix} 1 \\ 1 \\ \vdots \\ \vdots \\ 1 \\ 1 \end{bmatrix}_{nx1} \quad \beta = \begin{bmatrix} \beta_0 \\ \beta_0 \\ \vdots \\ \vdots \\ \beta_0 \\ \beta_0 \end{bmatrix}_{nx1} \quad \text{and} \quad \varepsilon = \begin{bmatrix} \varepsilon_{t_1} \\ \varepsilon_{t_2} \\ \vdots \\ \vdots \\ \varepsilon_{t_{n-1}} \\ \varepsilon_{t_n} \end{bmatrix}_{nx1}$$

From this presentation, (2) reduces to the form of $\hat{\beta}_0 = \frac{Y_1 + (1-\rho)\sum_{i=2}^{n-1} Y_i + Y_n}{2 + (1-\rho)(n-2)}$ with

$Var(\hat{\beta}_0) = \frac{1+\rho}{n(1-\rho)+2\rho}$ from (3). Table 1 shows the value of $\rho = e^{-\lambda/(n-1)}$, while Table 2 highlights the variance of $\hat{\beta}_0$ as both $n$ and $\lambda$ vary. Table 1 demonstrates that for any fixed $\lambda$, $\rho \to 1$ as $n \to \infty$ and for any fixed $n$, $\rho \to 0$ as $\lambda \to \infty$. This pattern for $\rho$ remains the same throughout all sections of this paper.

**Table 1.** Value of $\rho = e^{-\lambda/(n-1)}$ as $n$ and $\lambda$ vary.

| Sample size (*n*) | Lambda (*λ*) | | | | | |
|---|---|---|---|---|---|---|
| | 0.05 | 0.1 | 1 | 5 | 10 | 50 |
| 3 | 0.975310 | 0.951229 | 0.606531 | 0.082085 | 0.006738 | 1.389E-11 |
| 4 | 0.983471 | 0.967216 | 0.716531 | 0.188876 | 0.035674 | 5.778E-08 |
| 5 | 0.987578 | 0.975310 | 0.778801 | 0.286505 | 0.082085 | 3.727E-06 |
| 6 | 0.990050 | 0.980199 | 0.818731 | 0.367879 | 0.135335 | 4.540E-05 |
| 7 | 0.991701 | 0.983471 | 0.846482 | 0.434598 | 0.188876 | 2.404E-04 |
| 8 | 0.992883 | 0.985816 | 0.866878 | 0.489542 | 0.239651 | 7.905E-04 |
| 9 | 0.993769 | 0.987578 | 0.882497 | 0.535261 | 0.286505 | 0.001930 |
| 10 | 0.994460 | 0.988950 | 0.894839 | 0.573753 | 0.329193 | 0.003866 |
| 20 | 0.997372 | 0.994751 | 0.948730 | 0.768621 | 0.590778 | 0.071965 |
| 30 | 0.998277 | 0.996558 | 0.966105 | 0.841631 | 0.708343 | 0.178327 |
| 40 | 0.998719 | 0.997439 | 0.974685 | 0.879673 | 0.773824 | 0.277468 |
| 50 | 0.998980 | 0.997961 | 0.979799 | 0.902993 | 0.815396 | 0.360448 |

Table 2 suggests 1 as an upper bound for $Var(\hat{\beta}_0)$ and 0 as a lower bound (suggesting consistency). To investigate that, White (2001) derived the limiting variance of $Var(\hat{\beta}_0)$ as $\frac{2}{\lambda+2}$, which is independent of *n*. Since this variance is not equal to 0 as $n \to \infty$, $\hat{\beta}_0$ is therefore an inconsistent estimator for $\beta_0$ under an IA domain.

Using this limiting variance equation, it is easily noted that 0 is indeed the lower bound, while 1 is the upper bound as $\lambda \to \infty$ and $\lambda \to 0$, respectively. Figure 1 illustrates this pattern. Figure 1 also indicates that there is a diminishing return on the asymptotic value as $\lambda$ increases. This is seen in the knee in the curve in the figure. Beyond a certain $\lambda$ value, the decrease in asymptotic variance becomes negligible. In order to find this $\lambda$ value, the point at which the asymptotic variance decreases by <.0001 for a 0.1 increase in $\lambda$ is calculated. For this model, the maximum $\lambda$ that adds value is equal to 42.8.

**Table 2.** Variance of $Var(\hat{\beta}_0)$ as $n$ and $\lambda$ vary.

| Sample size ($n$) | Lambda ($\lambda$) | | | | | |
|---|---|---|---|---|---|---|
| | 0.05 | 0.1 | 1 | 5 | 10 | 50 |
| 3 | 0.975611 | 0.952390 | 0.671214 | 0.370842 | 0.336335 | 0.333333 |
| 4 | 0.975610 | 0.952385 | 0.668708 | 0.328215 | 0.263621 | 0.250000 |
| 5 | 0.975610 | 0.952383 | 0.667819 | 0.310714 | 0.227628 | 0.200001 |
| 6 | 0.975610 | 0.952382 | 0.667405 | 0.302061 | 0.207988 | 0.166679 |
| 7 | 0.975610 | 0.952382 | 0.667180 | 0.297202 | 0.196326 | 0.142916 |
| 8 | 0.975610 | 0.952382 | 0.667044 | 0.294216 | 0.188911 | 0.125173 |
| 9 | 0.975610 | 0.952382 | 0.666956 | 0.292254 | 0.183932 | 0.111493 |
| 10 | 0.975610 | 0.952381 | 0.666895 | 0.290899 | 0.180439 | 0.100698 |
| 20 | 0.975610 | 0.952381 | 0.666718 | 0.286889 | 0.169846 | 0.057310 |
| 30 | 0.975610 | 0.952381 | 0.666689 | 0.286219 | 0.168038 | 0.047120 |
| 40 | 0.975610 | 0.952381 | 0.666679 | 0.285994 | 0.167426 | 0.043368 |
| 50 | 0.975610 | 0.952381 | 0.666674 | 0.285891 | 0.167148 | 0.041606 |

To ascertain the rate at which $Var(\hat{\beta}_0) = \frac{1+\rho}{n(1-\rho)+2\rho} \to \frac{2}{2+\lambda}$, we take the derivative with respect to $n$. The derivative simplifies to $\frac{\frac{2\lambda\rho}{(n-1)}+\rho^2-1}{[n(1-\rho)+2\rho]^2}$, whose limit is 0 as $n \to \infty$. Table 3 highlights how quickly this occurs, while Figure 2 shows this for three select values of $\lambda$. Even though Figure 2, and others like it, has three subfigures with differing vertical scales, that is not the salient point of the figure. Their common trend of relatively quickly reaching the horizontal asymptote is.

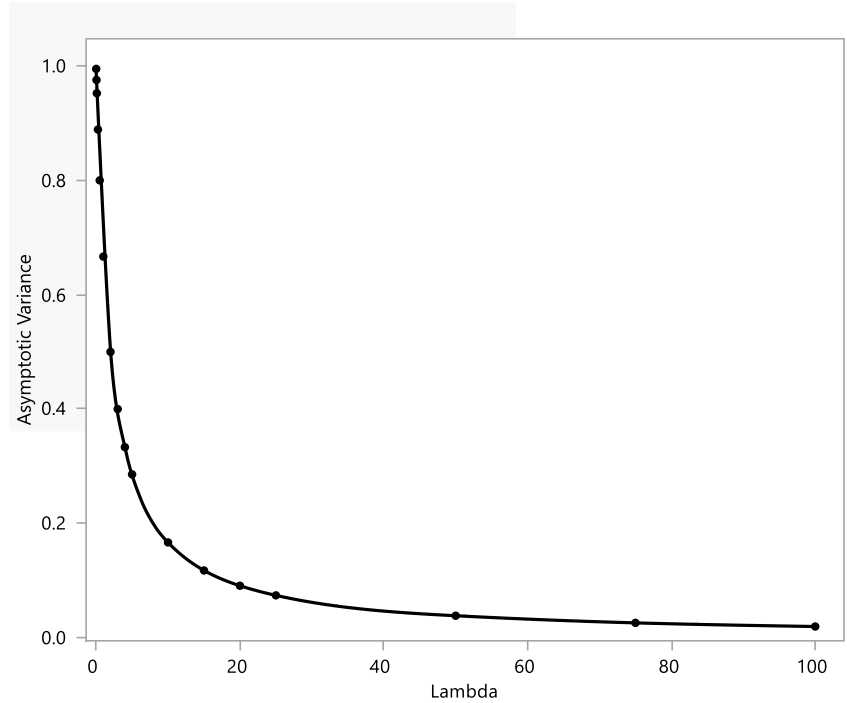

**Figure 1.** Asymptotic variance of $Var(\hat{\beta}_0)$ as a function of $\lambda$.

**Table 3.** Derivative evaluations of $Var(\hat{\beta}_0)$ as $n$ and $\lambda$ vary.

| Sample size ($n$) | 0.05 | 0.1 | Lambda ($\lambda$) 1 | 5 | 10 | 50 |
|---|---|---|---|---|---|---|
| 3 | -1.239E-06 | -9.444E-06 | -0.004467 | -0.068454 | -0.104086 | -0.111111 |
| 4 | -3.672E-07 | -2.799E-06 | -0.001350 | -0.025512 | -0.049299 | -0.062500 |
| 5 | -1.549E-07 | -1.181E-06 | -5.735E-04 | -0.011763 | -0.025791 | -0.039996 |
| 6 | -7.931E-08 | -6.046E-07 | -2.946E-04 | -0.006286 | -0.014778 | -0.027754 |
| 7 | -4.590E-08 | -3.499E-07 | -1.708E-04 | -0.003725 | -0.009128 | -0.020333 |
| 8 | -2.891E-08 | -2.204E-07 | -1.077E-04 | -0.002380 | -0.005988 | -0.015467 |
| 9 | -1.937E-08 | -1.476E-07 | -7.218E-05 | -0.001610 | -0.004122 | -0.012084 |
| 10 | -1.360E-08 | -1.037E-07 | -5.071E-05 | -0.001138 | -0.002950 | -0.009630 |
| 20 | -1.446E-09 | -1.102E-08 | -5.398E-06 | -1.233E-04 | -3.319E-04 | -0.001761 |
| 30 | -4.065E-10 | -3.099E-09 | -1.518E-06 | -3.478E-05 | -9.423E-05 | -5.649E-04 |
| 40 | -1.671E-10 | -1.274E-09 | -6.243E-07 | -1.432E-05 | -3.887E-05 | -2.438E-04 |
| 50 | -8.427E-11 | -6.425E-10 | -3.148E-07 | -7.222E-06 | -1.963E-05 | -1.258E-04 |

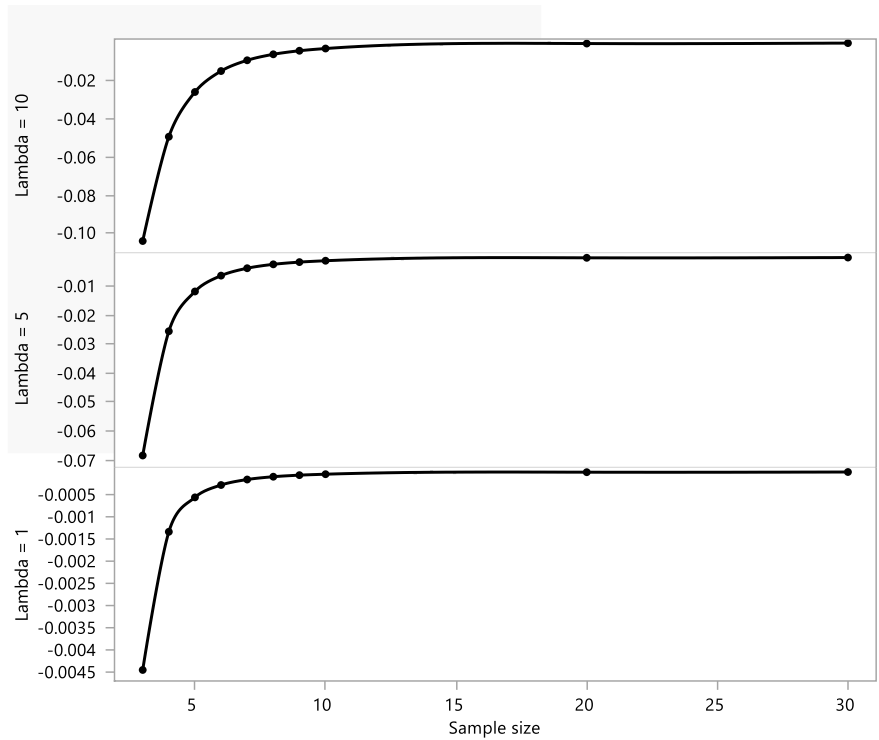

**Figure 2.** The derivative of $Var(\hat{\beta}_0)$ changing as a function of sample size *n* for $\lambda = 1, 5,$ and $10$.

Two interesting characteristics of the $Var(\hat{\beta}_0)$ appear from Table 3 and Figure 2. One, the variance is converging relatively quickly to the value $\frac{2}{2+\lambda}$ for small to moderate values of $\lambda$. Secondly, there appears to be a point of diminishing return as sample size increases for all values of $\lambda$. Parameter estimator inconsistency and point of diminishing return findings are contrary to the traditional thinking under an infinite domain. With consistent estimators, it follows that sampling infinitely will lead to the estimator variance equalling zero, i.e., sampling more is better. Under an IA domain, this line of thinking appears to be incorrect for the intercept only model. The next sections further investigate whether these conclusions hold for other models of (1).

# 4. SLOPE ONLY MODEL

The slope model presents the next simplest form of (1) whereby we assume the $y$-intercept is 0. Starting with (1) and assuming $n > 2$, let:

$$Y = \begin{bmatrix} Y_{t_1} \\ Y_{t_2} \\ \vdots \\ \vdots \\ Y_{t_{n-1}} \\ Y_{t_n} \end{bmatrix}_{nx1} \quad X = \begin{bmatrix} 0 \\ \frac{1}{n-1} \\ \frac{2}{n-1} \\ \vdots \\ \frac{n-2}{n-1} \\ 1 \end{bmatrix}_{nx1} \quad \boldsymbol{\beta} = \begin{bmatrix} \beta_1 \\ \beta_1 \\ \vdots \\ \vdots \\ \beta_1 \\ \beta_1 \end{bmatrix}_{nx1} \quad \text{and } \boldsymbol{\varepsilon} = \begin{bmatrix} \varepsilon_{t_1} \\ \varepsilon_{t_2} \\ \vdots \\ \vdots \\ \varepsilon_{t_{n-1}} \\ \varepsilon_{t_n} \end{bmatrix}_{nx1}$$

For $n = 2$, $X$ simply becomes $\begin{bmatrix} 0 \\ 1 \end{bmatrix}$. Our formulation of $X$ allows us to translate or standardize any line passing through the origin to a region confined to a line starting at (0,0) and ending at $(1, \beta_1)$. From this presentation, (2) reduces to

$$\hat{\beta}_1 = \frac{6\{-\rho Y_1 + (1-\rho)^2 \sum_{i=2}^{n-1}(i-1)Y_i + Y_n[2\rho + n(1-\rho) - 1]\}}{2n^2(1-2\rho+\rho^2) + n(8\rho-1) + \rho^2(6-7n)}, \text{ and from (3) the}$$

$$Var(\hat{\beta}_1) = \frac{6(1-\rho^2)(n-1)}{2n^2(1-2\rho+\rho^2) + n(8\rho-1) + \rho^2(6-7n)}.$$ Table 4 highlights the variance of $\hat{\beta}_1$ as both $n$ and $\lambda$ vary. Figure 3 illustrates this for select values of $\lambda$.

**Table 4.** Variance of $Var(\hat{\beta}_1)$ as $n$ and $\lambda$ vary.

| | | | Lambda ($\lambda$) | | | |
|---|---|---|---|---|---|---|
| Sample size ($n$) | 0.05 | 0.1 | 1 | 5 | 10 | 50 |
| 3 | 0.095162 | 0.181269 | 0.859514 | 0.849233 | 0.804292 | 0.800000 |
| 4 | 0.095162 | 0.181269 | 0.858241 | 0.777936 | 0.669011 | 0.642857 |
| 5 | 0.095162 | 0.181269 | 0.857769 | 0.745931 | 0.592750 | 0.533336 |
| 6 | 0.095162 | 0.181269 | 0.857546 | 0.729559 | 0.548762 | 0.454575 |
| 7 | 0.095162 | 0.181269 | 0.857424 | 0.720210 | 0.521888 | 0.395751 |
| 8 | 0.095162 | 0.181269 | 0.857350 | 0.714409 | 0.504517 | 0.350443 |
| 9 | 0.095162 | 0.181269 | 0.857302 | 0.710577 | 0.492732 | 0.314724 |
| 10 | 0.095162 | 0.181269 | 0.857268 | 0.707918 | 0.484406 | 0.286066 |
| 20 | 0.095162 | 0.181269 | 0.857171 | 0.700006 | 0.458873 | 0.166823 |
| 30 | 0.095162 | 0.181269 | 0.857155 | 0.698677 | 0.454473 | 0.137909 |
| 40 | 0.095162 | 0.181269 | 0.857150 | 0.698229 | 0.452981 | 0.127180 |
| 50 | 0.095162 | 0.181269 | 0.857147 | 0.698026 | 0.452303 | 0.122124 |

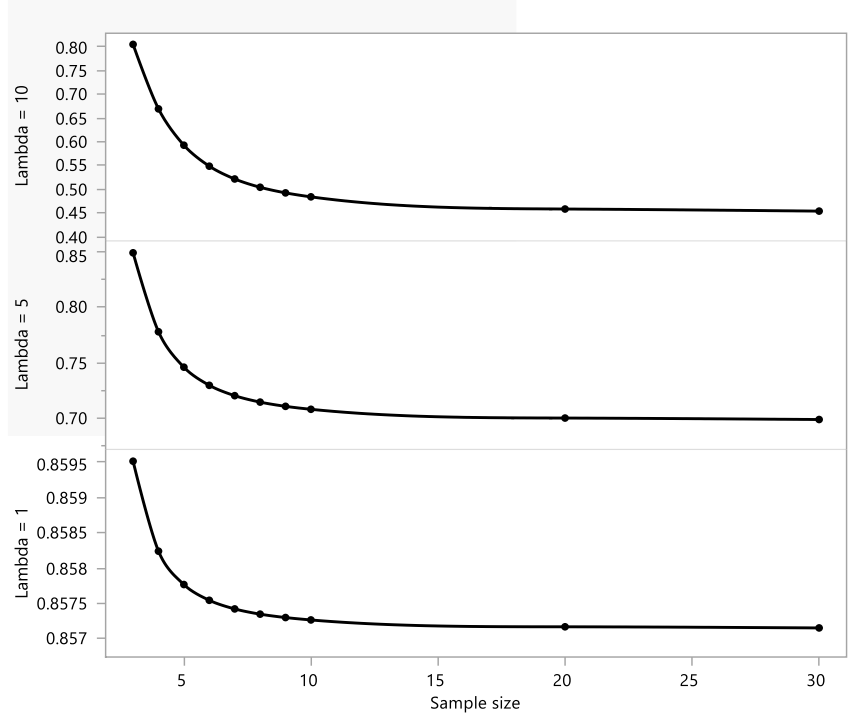

**Figure 3.** $Var(\hat{\beta}_1)$ versus sample size, $n$, for the slope only model for $\lambda = 1, 5,$ and $10$.

Table 4 suggests both an upper and lower bound for $Var(\hat{\beta}_1)$. Holding $n$ steady, $Var(\hat{\beta}_1) \to 0$ as $\lambda \to 0$ since $\rho \to 1$. However, the upper bound is a function of $n$, specifically $\frac{6(n-1)}{2n^2-n}$, as $\lambda \to \infty$ since $\rho \to 0$. To investigate the consistency of $Var(\hat{\beta}_1)$, we state and prove Theorem 1.

**Theorem 1:** $\lim_{n \to \infty} Var(\hat{\beta}_1) = \frac{12\lambda}{6+6\lambda+2\lambda^2} = \frac{2}{\frac{1}{\lambda}+1+\frac{\lambda}{3}}$

**Proof:** We work this in two parts: numerator followed by the denominator. In addition, we also utilize the Maclaurin series expression shown in (4) since both $\rho$ and $\rho^2$ involve $e$ via the expression $\rho = e^{-\lambda/(n-1)}$.

$$e^x = 1 + x + \frac{x^2}{2!} + \frac{x^3}{3!} + \cdots + \frac{x^n}{n!} + \cdots = \sum_{i=0}^{\infty} \frac{x^i}{i!} \tag{4}$$

Starting with $6(1 - \rho^2)(n - 1) = 6(n - n\rho^2 - 1 + \rho^2)$ when substituting we have: $6[n - ne^{-\frac{2\lambda}{n-1}} - 1 + e^{-\frac{2\lambda}{n-1}}]$. Substituting (4), we now have:

$$6[n - n(1 - \frac{2\lambda}{n-1} + \frac{4\lambda^2}{2(n-1)^2} + \cdots) - 1 + (1 - \frac{2\lambda}{n-1} + \frac{4\lambda^2}{2(n-1)^2} + \cdots)]$$

This is equivalent to:

$$6\left[n\frac{2\lambda}{n-1} - n\frac{4\lambda^2}{2(n-1)^2} + \cdots\right] - \frac{2\lambda}{n-1} + \frac{4\lambda^2}{2(n-1)^2} + \cdots$$

This simplifies to $12\lambda \left(\frac{n}{n-1}\right) + O\left(\frac{1}{n}\right)$, which converges to $12\lambda$ as $n \to \infty$.

For the denominator, we work $n^2(2 + 2\rho^2 - 4\rho) + n(8\rho - 1) + \rho^2(6 - 7n)$, separately in two pieces for ease of use. Starting with $n^2(2 + 2\rho^2 - 4\rho)$, we use (4) equalling:

$$n^2[2 + 2\left(1 - \frac{2\lambda}{n-1} + \frac{4\lambda^2}{2(n-1)^2} + \cdots\right) - 4\left(1 - \frac{\lambda}{n-1} + \frac{\lambda^2}{2(n-1)^2} + \cdots\right)]$$

This reduces to $\left(\frac{n}{n-1}\right)^2 2\lambda^2 + O\left(\frac{1}{n}\right)$ which converges to $2\lambda^2$ as $n \to \infty$. Focusing on the latter part of the denominator, we have $n(8\rho - 1) + \rho^2(6 - 7n)$. Via (4) we have:

$$n\left[8\left(1 - \frac{\lambda}{n-1} + \frac{\lambda^2}{2(n-1)^2} + \cdots\right) - 1\right] + 6\left(1 - \frac{2\lambda}{n-1} + \frac{4\lambda^2}{2(n-1)^2} + \cdots\right)$$

$$- 7n\left(1 - \frac{\lambda}{n-1} + \frac{4\lambda^2}{2(n-1)^2} + \cdots\right)$$

This reduces to $\left(\frac{n}{n-1}\right)6\lambda + 6 + O\left(\frac{1}{n}\right)$ and converges to $6\lambda + 6$ as $n \to \infty$. Putting the two pieces of the denominator together, we have that as $n \to \infty$, the denominator converges to $6 + 6\lambda + 2\lambda^2$. Combining these results, we reach:

$$\lim_{n \to \infty} Var(\hat{\beta}_1) \frac{12\lambda}{6+6\lambda+2\lambda^2} = \frac{2}{\frac{1}{\lambda}+1+\frac{\lambda}{3}} \qquad \text{Q.E.D.}$$

As shown in Theorem 1, $\hat{\beta}_1$ is not a consistent estimator for $\beta_1$ under an IA domain, however $\hat{\beta}_1$ is the UMVUE. These results are identical to what we concluded with the intercept only model. With this limiting variance equation, it is easily noted that 0 is indeed both the lower and upper bound as $\lambda \to 0$ and $\lambda \to \infty$. Figure 4 illustrates this pattern. The maximum value for this curve, 0.9282, is attained at $\lambda = \sqrt{3}$. Once again, this figure is used to calculate the maximum impactful $\lambda$ value with respect to asymptotic variance. The criterion is adjusted to only account for values of $\lambda$ that are greater than $\sqrt{3}$. Under this model, the maximum $\lambda$ value needed is 74.5.

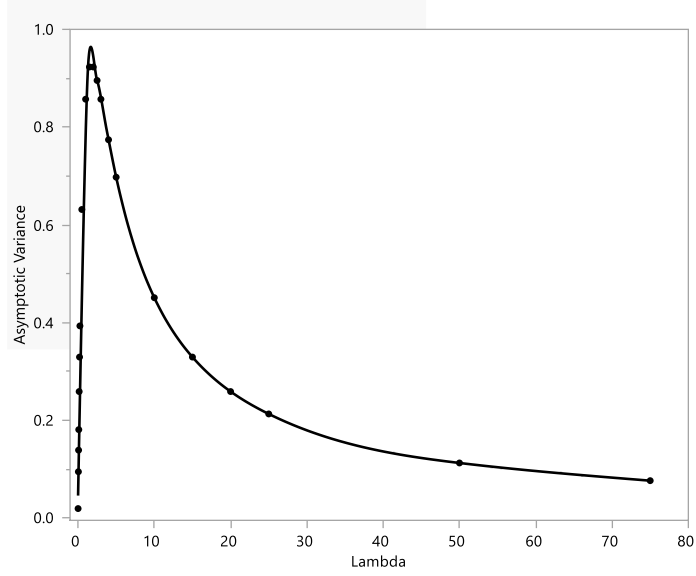

**Figure 4.** Asymptotic variance of $Var(\hat{\beta}_1)$ as a function of $\lambda$.

To establish the rate at which $Var(\hat{\beta}_1) \to \frac{2}{\frac{1}{\lambda}+1+\frac{\lambda}{3}}$, we take the derivative of $Var(\hat{\beta}_1)$ with respect to $n$. The derivative is expressed via the quotient rule as:

$\frac{E_3 E_2 - E_4 E_1}{E_3^2}$ where, $E_1 = 6(1-\rho^2)(n-1)$, $E_2 = 6 - 6\rho^2\left(1+\frac{2\lambda}{n-1}\right)$, $E_3 = 2n^2(1-2\rho+\rho^2) + n(8\rho-1) + \rho^2(6-7n)$, and $E_4 = (1-\rho)[4n(1-\rho) + 7\rho - 1] + \frac{2\lambda\rho(n-2)}{(n-1)^2}[2n(\rho-1) - 3\rho]$ whose limit approaches 0 as $n \to \infty$. Table 5 highlights how quickly this occurs, while Figure 5 shows this notionally for three select values of $\lambda$.

As with the intercept only model, we see the same two interesting characteristics. One, the variance is converging relatively quickly to $\frac{2}{\frac{1}{\lambda}+1+\frac{\lambda}{3}}$ for small to moderate values of $\lambda$. Secondly, there appears to be a point of diminishing return as sample size increases for all values of $\lambda$. The common thread for both the intercept only and slope only models are parameter estimator inconsistency in the IA domain and a point of diminishing return with respect to collecting more data. We now turn to our last model whereby we allow for two parameters simultaneously varying.

**Table 5.** Derivative evaluations of $Var(\hat{\beta}_1)$ as $n$ and $\lambda$ vary.

|  |  |  | Lambda ($\lambda$) |  |  |  |
|---|---|---|---|---|---|---|
| Sample size ($n$) | 0.05 | 0.1 | 1 | 5 | 10 | 50 |
| 3 | -3.537E-09 | -1.026E-07 | -0.002196 | -0.106494 | -0.174382 | -0.186667 |
| 4 | -1.113E-09 | -3.230E-08 | -7.082E-04 | -0.045541 | -0.100439 | -0.130102 |
| 5 | -4.790E-10 | -1.390E-08 | -3.075E-04 | -0.022015 | -0.056693 | -0.091844 |
| 6 | -2.475E-10 | -7.184E-09 | -1.595E-04 | -0.012023 | -0.033704 | -0.067436 |
| 7 | -1.440E-10 | -4.178E-09 | -9.298E-05 | -0.007209 | -0.021250 | -0.051252 |
| 8 | -9.093E-11 | -2.639E-09 | -5.880E-05 | -0.004639 | -0.014116 | -0.040001 |
| 9 | -6.103E-11 | -1.771E-09 | -3.950E-05 | -0.003152 | -0.009797 | -0.031849 |
| 10 | -4.292E-11 | -1.246E-09 | -2.780E-05 | -0.002235 | -0.007052 | -0.025749 |
| 20 | -4.579E-12 | -1.329E-10 | -2.971E-06 | -2.444E-04 | -8.065E-04 | -0.004967 |
| 30 | -1.289E-12 | -3.741E-11 | -8.362E-07 | -6.905E-05 | -2.296E-04 | -0.001612 |
| 40 | -5.302E-13 | -1.538E-11 | -3.439E-07 | -2.844E-05 | -9.481E-05 | -6.988E-04 |
| 50 | -2.673E-13 | -7.757E-12 | -1.734E-07 | -1.435E-05 | -4.789E-05 | -3.611E-04 |

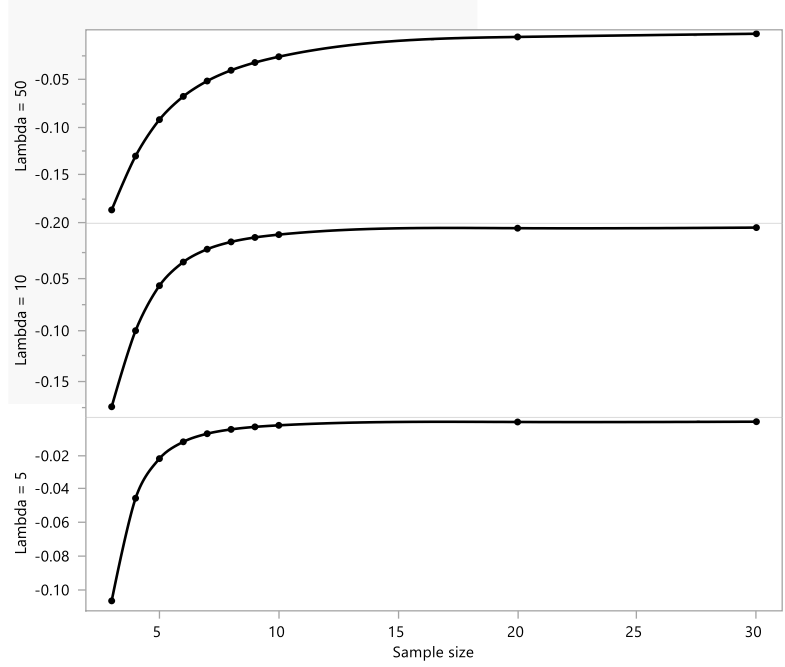

**Figure 5.** The derivative of $Var(\hat{\beta}_1)$ changing as a function of sample size *n* for $\lambda = 5, 10,$ and $50$.

## 5. INTERCEPT and SLOPE MODEL

The intercept and slope model presents a more complex form of (1) that contains two parameters. Starting with (1) and assuming $n > 2$, we have the following:

$$\boldsymbol{Y} = \begin{bmatrix} Y_{t_1} \\ Y_{t_2} \\ \vdots \\ \vdots \\ Y_{t_{n-1}} \\ Y_{t_n} \end{bmatrix}_{nx1} \quad \boldsymbol{X} = \begin{bmatrix} 1 & 0 \\ 1 & \frac{1}{n-1} \\ \vdots & \vdots \\ 1 & \frac{n-2}{n-1} \\ 1 & 1 \end{bmatrix}_{nx2} \quad \boldsymbol{\beta} = \begin{bmatrix} \beta_0 & \beta_1 \\ \beta_0 & \beta_1 \\ \vdots & \vdots \\ \vdots & \vdots \\ \beta_0 & \beta_1 \\ \beta_0 & \beta_1 \end{bmatrix}_{nx2} \quad \text{and } \boldsymbol{\varepsilon} = \begin{bmatrix} \varepsilon_{t_1} \\ \varepsilon_{t_2} \\ \vdots \\ \vdots \\ \varepsilon_{t_{n-1}} \\ \varepsilon_{t_n} \end{bmatrix}_{nx1}$$

Under an increasing domain, it is known the estimators of $\boldsymbol{\beta}$ display a dependency, specifically the general covariance equals $-\bar{X}Var(\hat{\beta}_1)$. Besides investigating the individual variance of the estimators where both the intercept and slope can vary, we will also investigate the covariance of $\widehat{\boldsymbol{\beta}}$ under an IA.

*5.1 Intercept*

The first parameter of interest is the intercept. From (2) the form of $\hat{\beta}_0$ is relatively cumbersome and entails the determinant of a 2 x 2 matrix. To facilitate subsequent sections, we will express this determinant in various elements. It can be shown that $X'\Sigma^{-1}X$ has the form, $\frac{1}{1-\rho^2}\begin{bmatrix} M_1 & M_2 \\ M_2 & M_3 \end{bmatrix}$, where $M_1 = 2(1-\rho) + (n-2)(1-\rho)^2$, $M_2 = \frac{(1-\rho)^2(n-2)}{2} + (1-\rho)$, and $M_3 = \frac{(1-\rho)^2}{n-1}\left[\frac{(n-2)(2n-3)}{6}\right] - \frac{\rho(n-2)}{n-1} + 1$. Therefore, $(X'\Sigma^{-1}X)^{-1}$ equals $\frac{(1-\rho^2)}{det}\begin{bmatrix} M_3 & -M_2 \\ -M_2 & M_1 \end{bmatrix}$, where $det = M_1 M_3 - (M_2)^2$. $X'\Sigma^{-1}Y$ has the form $\begin{bmatrix} V_1 \\ V_2 \end{bmatrix}$, where $V_1 = \frac{1}{1+\rho}[Y_1 + (1-\rho)\sum_{i=2}^{n-1} Y_i + Y_n]$ and $V_2 = \left(\frac{1}{1-\rho^2}\right)\left(\frac{1}{n-1}\right)\{-\rho Y_1 + (1-\rho)^2 \sum_{i=2}^{n-1}(i-1)Y_i + [-\rho(n-2) + (n-1)]Y_n\}$.

Putting this altogether, we have that $\hat{\beta}_0 = \frac{(1-\rho^2)}{det}(M_3 V_1 - M_2 V_2)$.

It can be shown that $Var(\hat{\beta}_0) = \frac{2(1+\rho)[2n^2(1-2\rho+\rho^2)+n(8\rho-1-7\rho^2)+6\rho^2]}{[n(1-\rho)+2\rho][n^2(1-2\rho+\rho^2)+n(4\rho+1-5\rho^2)+6\rho(1+\rho)]}$ from (3). Table 6 highlights the variance of $\hat{\beta}_0$ as both $n$ and $\lambda$ vary while Figure 6 displays the information graphically for select $\lambda$ values.

**Table 6.** $Var(\hat{\beta}_0)$ for the intercept and slope model with respect to $n$ and $\lambda$.

| | | | Lambda ($\lambda$) | | | |
|---|---|---|---|---|---|---|
| Sample size (*n*) | 0.05 | 0.1 | 1 | 5 | 10 | 50 |
| 3 | 0.999996 | 0.999972 | 0.997564 | 0.867473 | 0.836312 | 0.833333 |
| 4 | 0.999996 | 0.999966 | 0.997098 | 0.805536 | 0.721151 | 0.700000 |
| 5 | 0.999995 | 0.999965 | 0.996933 | 0.777126 | 0.652244 | 0.600002 |
| 6 | 0.999995 | 0.999964 | 0.996857 | 0.762429 | 0.611026 | 0.523838 |
| 7 | 0.999995 | 0.999963 | 0.996816 | 0.753985 | 0.585294 | 0.464433 |
| 8 | 0.999995 | 0.999963 | 0.996791 | 0.748727 | 0.568436 | 0.417128 |
| 9 | 0.999995 | 0.999963 | 0.996774 | 0.745245 | 0.556897 | 0.378846 |
| 10 | 0.999995 | 0.999963 | 0.996763 | 0.742826 | 0.548696 | 0.347483 |
| 20 | 0.999995 | 0.999962 | 0.996731 | 0.735607 | 0.523289 | 0.210483 |
| 30 | 0.999995 | 0.999962 | 0.996725 | 0.734393 | 0.518871 | 0.175618 |
| 40 | 0.999995 | 0.999962 | 0.996724 | 0.733983 | 0.517370 | 0.162511 |
| 50 | 0.999995 | 0.999962 | 0.996723 | 0.733797 | 0.516687 | 0.156303 |

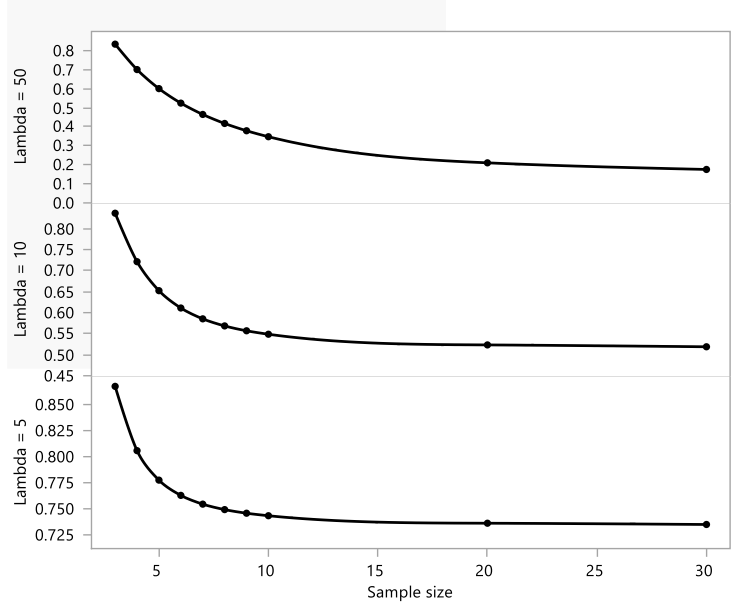

**Figure 6.** $Var(\hat{\beta}_0)$ versus sample size, $n$, for the intercept and slope model for $\lambda = 5, 10,$ and $50$.

Table 6 suggests 1 is an upper bound for $Var(\hat{\beta}_0)$, with a possible lower bound of 0. To investigate that, the limiting variance of $Var(\hat{\beta}_0)$ is derived.

**Theorem 2:** $\lim\limits_{n \to \infty} Var(\hat{\beta}_0) = \dfrac{8(\lambda^2 + 3\lambda + 3)}{(\lambda+2)(\lambda^2 + 6\lambda + 12)}$

**Proof:** As with Theorem 1, we work this in two parts, the numerator followed by the denominator, in conjunction with (4). The numerator, $2(1+\rho)[2n^2(1-2\rho+\rho^2) + n(8\rho - 1 - 7\rho^2) + 6\rho^2]$ can be expressed as: $2\left[1 + \left(1 - \frac{\lambda}{n-1} + \cdots\right)\right]\left\{2n^2\left[1 - 2\left(1 - \frac{\lambda}{n-1} + \frac{\lambda^2}{2(n-1)^2} + \cdots\right) + \left(1 - \frac{2\lambda}{n-1} + \frac{4\lambda^2}{2(n-1)^2} + \cdots\right)\right] + n\left[8\left(1 - \frac{\lambda}{n-1} + \cdots\right) - 1 - 7\left(1 - \frac{2\lambda}{n-1} + \frac{4\lambda^2}{2(n-1)^2} + \cdots\right)\right] + 6\left(1 - \frac{2\lambda}{n-1} + \cdots\right)\right\}$ This simplifies to: $2\left(2 + O\left(\frac{1}{n}\right)\right)\left[2\lambda^2\left(\frac{n}{n-1}\right)^2 + 6\lambda\left(\frac{n}{n-1}\right) + 6 + O\left(\frac{1}{n}\right)\right]$. As $n \to \infty$, the numerator converges to $4(2\lambda^2 + 6\lambda + 6)$, which equals $8(\lambda^2 + 3\lambda + 3)$.

The denominator follows a similar logic. Specifically, $[n(1-\rho) + 2\rho][n^2(1 - 2\rho + \rho^2) + n(4\rho + 1 - 5\rho^2) + 6\rho(1+\rho)]$ can be expressed as:

$\left[n\left(1-\left(1-\frac{\lambda}{n-1}+\cdots\right)\right)+2\left(1-\frac{\lambda}{n-1}+\cdots\right)\right]\left\{n^2\left[1-2\left(1-\frac{\lambda}{n-1}+\frac{\lambda^2}{2(n-1)^2}+\cdots\right)+\right.\right.$

$\left(1-\frac{2\lambda}{n-1}+\frac{4\lambda^2}{2(n-1)^2}+\cdots\right)\right]+n\left[4\left(1-\frac{\lambda}{n-1}+\cdots\right)+1-5\left(1-\frac{2\lambda}{n-1}+\frac{4\lambda^2}{2(n-1)^2}+\cdots\right)\right]+$

$6\left(1-\frac{\lambda}{n-1}+\cdots\right)\left[1+\left(1-\frac{\lambda}{n-1}+\cdots\right)\right]\}$. This simplifies to: $\left[\lambda\left(\frac{n}{n-1}\right)+2+\right.$

$\left.O\left(\frac{1}{n}\right)\right]\left[\lambda^2\left(\frac{n}{n-1}\right)^2+6\lambda\left(\frac{n}{n-1}\right)+12+O\left(\frac{1}{n}\right)\right]$. As $n\to\infty$, the denominator converges to

$(\lambda+2)(\lambda^2+6\lambda+12)$. Putting this together, we have:

$$\lim_{n\to\infty}Var(\hat{\beta}_0)=\frac{8(\lambda^2+3\lambda+3)}{(\lambda+2)(\lambda^2+6\lambda+12)} \qquad \text{Q.E.D.}$$

Since this variance is not equal to zero as $n\to\infty$, $\hat{\beta}_0$ is an inconsistent estimator for $\beta_0$ under this model in an IA domain. With this limiting variance equation, it is noted that 0 is indeed the lower bound, while 1 is the upper bound as $\lambda\to\infty$ and $\lambda\to0$, respectively. Figure 7 illustrates this pattern. Again, this figure is used to calculate the maximum impactful $\lambda$ value. Under this model, the maximum $\lambda$ value needed is 84.5.

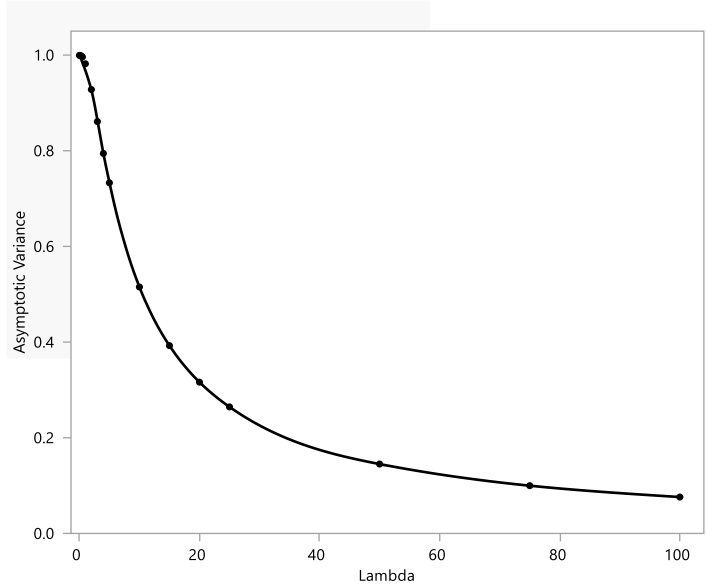

**Figure 7.** Asymptotic variance of $Var(\hat{\beta}_0)$, for the intercept and slope model, as a function of $\lambda$.

To determine the rate at which $Var(\hat{\beta}_0) \to \frac{8(\lambda^2+3\lambda+3)}{(\lambda+2)(\lambda^2+6\lambda+12)}$, we evaluate the derivative of $Var(\hat{\beta}_0)$ with respect to $n$. The derivative is expressed via the quotient rule as: $\frac{E_3 E_2 - E_4 E_1}{E_3^2}$ where, $E_1 = 2(1+\rho)[2n^2(1-2\rho+\rho^2) + n(8\rho - 1 - 7\rho^2) + 6\rho^2]$, $E_2 = \frac{2\lambda\rho}{(n-1)^2}[2n^2(1-2\rho+\rho^2) + n(8\rho - 1 - 7\rho^2) + 6\rho^2] + 2(1+\rho)[4n(1-2\rho+\rho^2) + 2n^2\left[\frac{-2\lambda\rho}{(n-1)^2} + \frac{2\lambda\rho^2}{(n-1)^2}\right] + (8\rho - 1 - 7\rho^2) + n\left[\frac{8\lambda\rho}{(n-1)^2} - \frac{14\lambda\rho^2}{(n-1)^2}\right] + \frac{12\lambda\rho^2}{(n-1)^2}]$, $E_3 = [n(1-\rho) + 2\rho][n^2(1-2\rho+\rho^2) + n(4\rho+1-5\rho^2) + 6\rho(1+\rho)]$, and $E_4 = [(1-\rho) - \frac{n\lambda\rho}{(n-1)^2} + \frac{2\lambda\rho}{(n-1)^2}][n^2(1-2\rho+\rho^2) + n(4\rho+1-5\rho^2 + 6\rho(1+\rho)] + [n(1-\rho) + 2\rho]\left[2n(1-2\rho+\rho^2) + n^2\left[\frac{-2\lambda\rho}{(n-1)^2} + \frac{2\lambda\rho^2}{(n-1)^2}\right] + (4\rho+1-5\rho^2) + n\left[\frac{4\lambda\rho}{(n-1)^2} - \frac{10\lambda\rho^2}{(n-1)^2}\right] + \frac{6\lambda\rho(1+\rho)}{(n-1)^2} + \frac{6\lambda\rho^2}{(n-1)^2}\right]$, whose limit quickly approaches 0 as $n \to \infty$. Table 7 highlights how quickly this occurs, while Figure 8 shows this notionally for three select values of $\lambda$.

**Table 7.** Derivative evaluations of $Var(\hat{\beta}_0)$, intercept and slope model, as $n$ and $\lambda$ vary.

| Sample size ($n$) | | | Lambda ($\lambda$) | | | |
|---|---|---|---|---|---|---|
| | 0.05 | 0.1 | 1 | 5 | 10 | 50 |
| 3 | -1.24E-06 | -9.45E-06 | -0.004664 | -0.091364 | -0.143621 | -0.152778 |
| 4 | -3.67E-07 | -2.80E-06 | -0.001433 | -0.040134 | -0.088705 | -0.115000 |
| 5 | -1.55E-07 | -1.18E-06 | -0.000612 | -0.019687 | -0.052415 | -0.086659 |
| 6 | -7.93E-08 | -6.05E-07 | -0.000315 | -0.010835 | -0.032008 | -0.066837 |
| 7 | -4.59E-08 | -3.50E-07 | -0.000183 | -0.006526 | -0.020514 | -0.052736 |
| 8 | -2.89E-08 | -2.21E-07 | -0.000115 | -0.004211 | -0.013771 | -0.042377 |
| 9 | -1.94E-08 | -1.48E-07 | -7.74E-05 | -0.002866 | -0.009626 | -0.034531 |
| 10 | -1.36E-08 | -1.04E-07 | -5.44E-05 | -0.002035 | -0.006963 | -0.028443 |
| 20 | -1.45E-09 | -1.10E-08 | -5.80E-06 | -0.000223 | -0.000809 | -0.005933 |
| 30 | -4.07E-10 | -3.10E-09 | -1.63E-06 | -6.32E-05 | -0.000231 | -0.001963 |
| 40 | -1.67E-10 | -1.28E-09 | -6.71E-07 | -2.60E-05 | -9.54E-05 | -0.000857 |
| 50 | -8.43E-11 | -6.43E-10 | -3.38E-07 | -1.31E-05 | -4.82E-05 | -0.000444 |

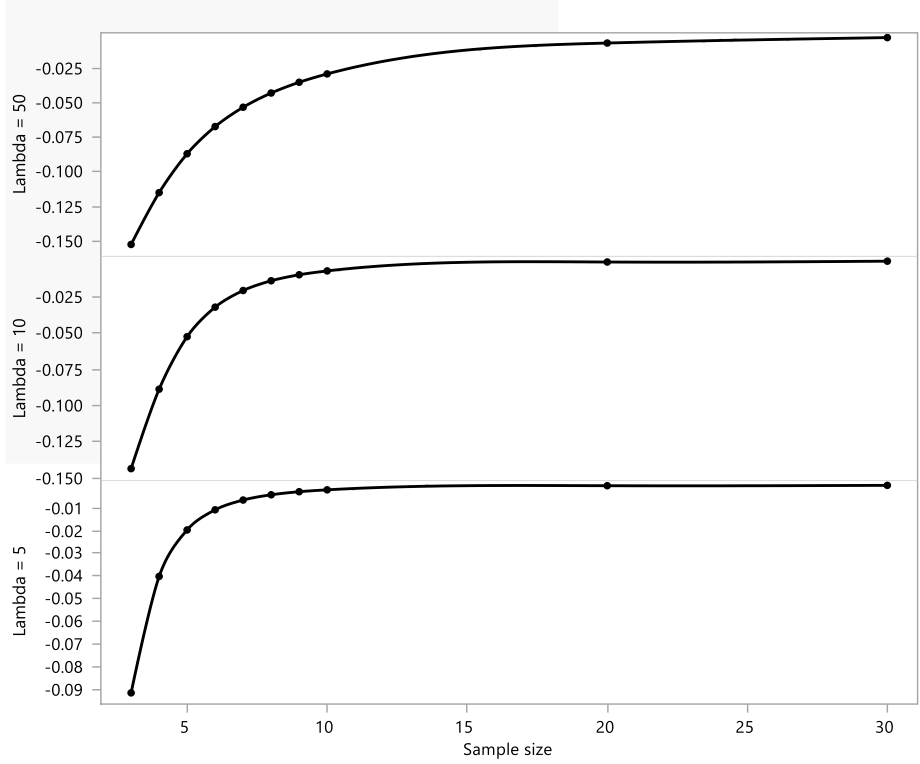

**Figure 8.** The derivative of $Var(\hat{\beta}_0)$ changing as a function of sample size $n$ (intercept and slope model) for $\lambda = 5, 10,$ and $50$.

As with the previous models, we see the same two interesting characteristics. One, the variance is converging relatively quickly to $\frac{8(\lambda^2+3\lambda+3)}{(\lambda+2)(\lambda^2+6\lambda+12)}$ for small to moderate values of $\lambda$. Secondly, there appears to be a point of diminishing return as sample size increases for all values of $\lambda$. The common thread continues to be estimator inconsistency in the IA domain and a point of diminishing return with respect to collecting more data.

*5.2 Slope*

For the slope, (2) reduces to $\hat{\beta}_1 = \frac{(1-\rho^2)}{det}(-M_2 V_1 + M_1 V_2)$. using the terms defined in subsection 5.1. From (3), $Var(\hat{\beta}_1) = \frac{12(1-\rho^2)(n-1)}{n^2(1+\rho^2-2\rho)+n(1+4\rho-5\rho^2)+6\rho(1+\rho)}$. Table 8 highlights the variance of $\hat{\beta}_1$ as both $n$ and $\lambda$ vary and Figure 9 shows this graphically.

**Table 8.** $Var(\hat{\beta}_1)$ for the intercept and slope model with respect to $n$ and $\lambda$.

| Sample size ($n$) | Lambda ($\lambda$) | | | | | |
|---|---|---|---|---|---|---|
| | 0.05 | 0.1 | 1 | 5 | 10 | 50 |
| 3 | 0.097541 | 0.190325 | 1.264241 | 1.986524 | 1.999909 | 2.000000 |
| 4 | 0.097541 | 0.190325 | 1.263713 | 1.909284 | 1.830120 | 1.800000 |
| 5 | 0.097541 | 0.190325 | 1.263485 | 1.865650 | 1.698465 | 1.600005 |
| 6 | 0.097541 | 0.190325 | 1.263372 | 1.841470 | 1.612151 | 1.428636 |
| 7 | 0.097541 | 0.190325 | 1.263308 | 1.827130 | 1.555872 | 1.286067 |
| 8 | 0.097541 | 0.190325 | 1.263269 | 1.818044 | 1.518102 | 1.167820 |
| 9 | 0.097541 | 0.190325 | 1.263243 | 1.811964 | 1.491863 | 1.069413 |
| 10 | 0.097541 | 0.190325 | 1.263226 | 1.807709 | 1.473030 | 0.987139 |
| 20 | 0.097541 | 0.190325 | 1.263173 | 1.794875 | 1.413771 | 0.612691 |
| 30 | 0.097541 | 0.190325 | 1.263164 | 1.792695 | 1.403331 | 0.513992 |
| 40 | 0.097541 | 0.190325 | 1.263162 | 1.791959 | 1.399776 | 0.476572 |
| 50 | 0.097541 | 0.190325 | 1.263160 | 1.791624 | 1.398158 | 0.458790 |

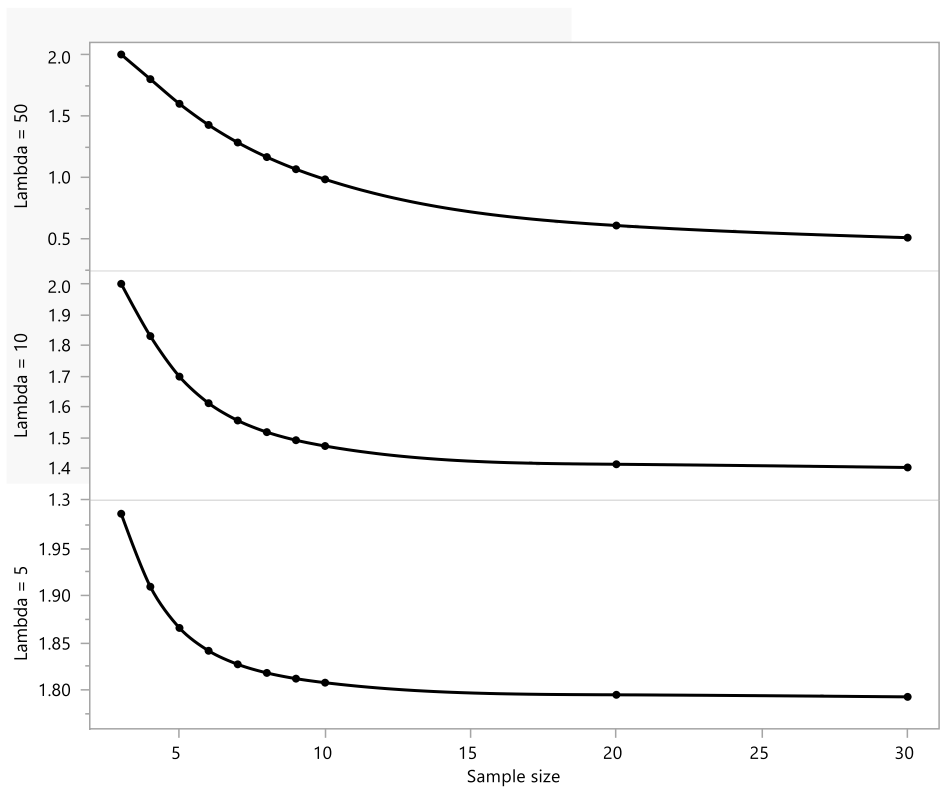

**Figure 9.** $Var(\hat{\beta}_1)$ versus sample size, $n$, for the intercept and slope model for $\lambda = 5, 10,$ and $50$.

Table 8 suggests both an upper and lower bound for $Var(\hat{\beta}_1)$. Holding $n$ steady, $Var(\hat{\beta}_1) \to 0$ as $\lambda \to 0$ since $\rho \to 1$. However, the upper bound is a function of $n$, specifically $\frac{12(n-1)}{n^2+n}$, as $\lambda \to \infty$ since $\rho \to 0$. To investigate the consistency or inconsistency of $Var(\hat{\beta}_1)$, we take the limit as $n \to \infty$.

**Theorem 3:** $\lim_{n \to \infty} Var(\hat{\beta}_1) = \frac{24\lambda}{12+6\lambda+\lambda^2} = \frac{4}{\frac{2}{\lambda}+1+\frac{\lambda}{6}}$

**Proof:** We work this in two parts: numerator followed by the denominator. In addition, we also utilize the Maclaurin series expression shown in (4). Starting with $12(1-\rho^2)(n-1) = 12(n - n\rho^2 - 1 + \rho^2)$ and substituting (4), we have:

$$12[n - n(1 - \frac{2\lambda}{n-1} + \frac{4\lambda^2}{2(n-1)^2} + \cdots) - 1 + (1 - \frac{2\lambda}{n-1} + \frac{4\lambda^2}{2(n-1)^2} + \cdots)].$$

This is equivalent to $12\left[n\frac{2\lambda}{n-1} - n\frac{4\lambda^2}{2(n-1)^2} + \cdots\right] - \frac{2\lambda}{n-1} + \frac{4\lambda^2}{2(n-1)^2} + \cdots$, and simplifies to $24\lambda\left(\frac{n}{n-1}\right) + O\left(\frac{1}{n}\right)$, which converges to $24\lambda$ as $n \to \infty$.

For the denominator, we work $n^2(1 + \rho^2 - 2\rho) + n(1 + 4\rho - 5\rho^2) + 6\rho(1 + \rho)$, separately in two pieces for ease of use. Starting with $n^2(1 + \rho^2 - 2\rho)$, we use (4) equalling:

$$n^2[1 + \left(1 - \frac{2\lambda}{n-1} + \frac{4\lambda^2}{2(n-1)^2} + \cdots\right) - 2\left(1 - \frac{\lambda}{n-1} + \frac{\lambda^2}{2(n-1)^2} + \cdots\right)]$$

This reduces to $\left(\frac{n}{n-1}\right)^2 \lambda^2 + O\left(\frac{1}{n}\right)$ which converges to $\lambda^2$ as $n \to \infty$. Focusing on the latter part of the denominator, we have $n(1 + 4\rho - 5\rho^2) + 6\rho(1 + \rho)$. Via (4) we have:

$$n\left[1 + 4\left(1 - \frac{\lambda}{n-1} + \frac{\lambda^2}{2(n-1)^2} + \cdots\right) - 5\left(1 - \frac{2\lambda}{n-1} + \frac{4\lambda^2}{2(n-1)^2} + \cdots\right)\right]$$

$$+ 6\left(1 - \frac{2\lambda}{n-1} + \frac{4\lambda^2}{2(n-1)^2} + \cdots\right)\left(2 - \frac{2\lambda}{n-1} + \frac{4\lambda^2}{2(n-1)^2} + \cdots\right)$$

This reduces to $\left(\frac{n}{n-1}\right)6\lambda + 12 + O\left(\frac{1}{n}\right)$ and converges to $6\lambda + 12$ as $n \to \infty$. Putting the two pieces of the denominator together, we have that as $n \to \infty$, the denominator converges to $12 + 6\lambda + \lambda^2$. Combining these, we are left with:

$$\lim_{n\to\infty} Var(\hat{\beta}_1) = \frac{24\lambda}{12+6\lambda+\lambda^2} = \frac{4}{\frac{2}{\lambda}+1+\frac{\lambda}{6}} \quad \text{Q.E.D.}$$

As shown in Theorem 3, $\hat{\beta}_1$ is not a consistent estimator for $\beta_1$ under an IA domain. These results are identical to what we concluded previously. With this limiting variance equation, it is easily noted that 0 is indeed both the lower and upper bound as $\lambda \to 0$ and $\lambda \to \infty$. Figure 10 illustrates this pattern. The maximum value for this curve, 1.8564, is attained at $\lambda = \sqrt{12}$. Once more, this figure is used to observe the maximum impactful $\lambda$ value. The criterion is adjusted to only account for values of $\lambda$ that are greater than $\sqrt{12}$. Under this model, the maximum $\lambda$ value needed is 148.9.

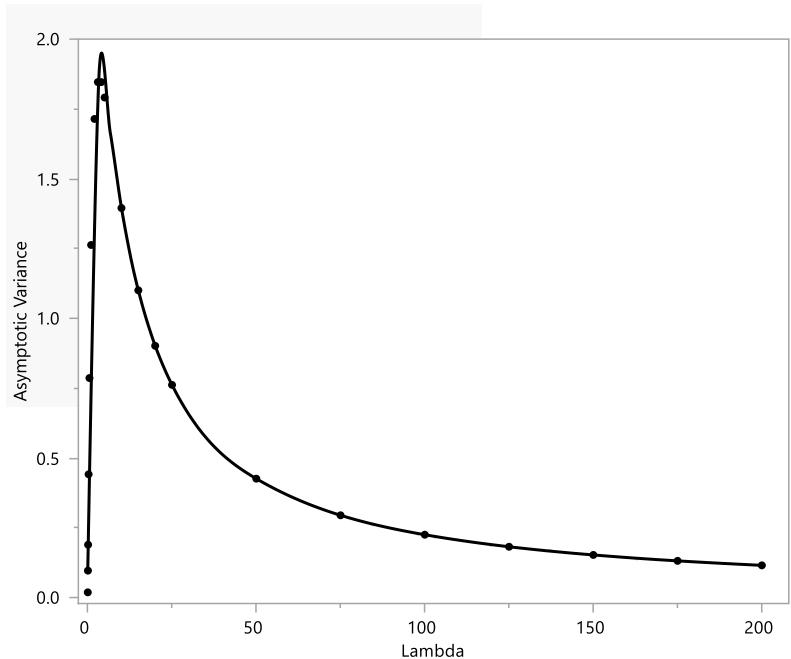

**Figure 10.** Asymptotic variance of $Var(\hat{\beta}_1)$, for the intercept and slope model, as a function of $\lambda$.

To ascertain the rate at which $Var(\hat{\beta}_1) \to \frac{4}{\frac{2}{\lambda}+1+\frac{\lambda}{6}}$, we take the derivative of

$Var(\hat{\beta}_1)$ with respect to $n$. The derivative is expressed via the quotient rule as:

$\frac{E_3 E_2 - E_4 E_1}{E_3^2}$ where, $E_1 = 12(1-\rho^2)(n-1)$, $E_2 = \frac{-24\lambda\rho^2}{n-1} + 12(1-\rho^2)$, $E_3 = n^2(1+\rho^2-2\rho) + n(1+4\rho-5\rho^2) + 6\rho(1+\rho)$ and $E_4 = 2n(1-2\rho+\rho^2) + n^2\left[\frac{-2\lambda\rho}{(n-1)^2} + \frac{2\lambda\rho^2}{(n-1)^2}\right] + (1+4\rho-5\rho^2) + n\left[\frac{4\lambda\rho}{(n-1)^2} - \frac{10\lambda\rho^2}{(n-1)^2}\right] + \frac{6\lambda\rho(1+2\rho)}{(n-1)^2}$, whose limit quickly approaches 0 as $n \to \infty$. Table 9 highlights how quickly this occurs, while Figure 11 shows this notationally for two select values of $\lambda$.

Again, the same two characteristics, the variance is converging relatively quickly to $\frac{4}{\frac{2}{\lambda}+1+\frac{\lambda}{6}}$ for small to moderate values of $\lambda$ and there appears to be a point of diminishing return as sample size increases for all values of $\lambda$, are noticeable. The slope parameter estimator is inconsistent in the IA domain for the intercept and slope model as well. We now examine the covariance under this model.

**Table 9.** Derivative evaluations of $Var(\hat{\beta}_1)$, intercept and slope model, as $n$ and $\lambda$ vary.

| | | | Lambda ($\lambda$) | | | |
|---|---|---|---|---|---|---|
| Sample size ($n$) | 0.05 | 0.1 | 1 | 5 | 10 | 50 |
| 3 | -6.20E-10 | -1.90E-08 | -0.000790 | -0.091640 | -0.158137 | -0.166667 |
| 4 | -2.50E-10 | -7.70E-09 | -0.000330 | -0.058488 | -0.157623 | -0.210000 |
| 5 | -1.20E-10 | -3.50E-09 | -0.000154 | -0.031696 | -0.106494 | -0.186651 |
| 6 | -6.20E-11 | -1.90E-09 | -8.20E-05 | -0.018197 | -0.068919 | -0.156329 |
| 7 | -3.70E-11 | -1.10E-09 | -4.90E-05 | -0.011203 | -0.045541 | -0.129609 |
| 8 | -2.30E-11 | -7.10E-10 | -3.10E-05 | -0.007323 | -0.031131 | -0.107641 |
| 9 | -1.60E-11 | -4.80E-10 | -2.10E-05 | -0.005026 | -0.022015 | -0.089788 |
| 10 | -1.10E-11 | -3.40E-10 | -1.50E-05 | -0.003589 | -0.016053 | -0.075254 |
| 20 | -1.20E-12 | -3.70E-11 | -1.60E-06 | -0.000400 | -0.001908 | -0.016689 |
| 30 | -3.40E-13 | -1.00E-11 | -4.50E-07 | -0.000113 | -0.000547 | -0.005590 |
| 40 | -1.40E-13 | -4.30E-12 | -1.90E-07 | -0.000047 | -0.000226 | -0.002451 |
| 50 | -7.00E-14 | -2.10E-12 | -9.40E-08 | -0.000024 | -0.000114 | -0.001274 |

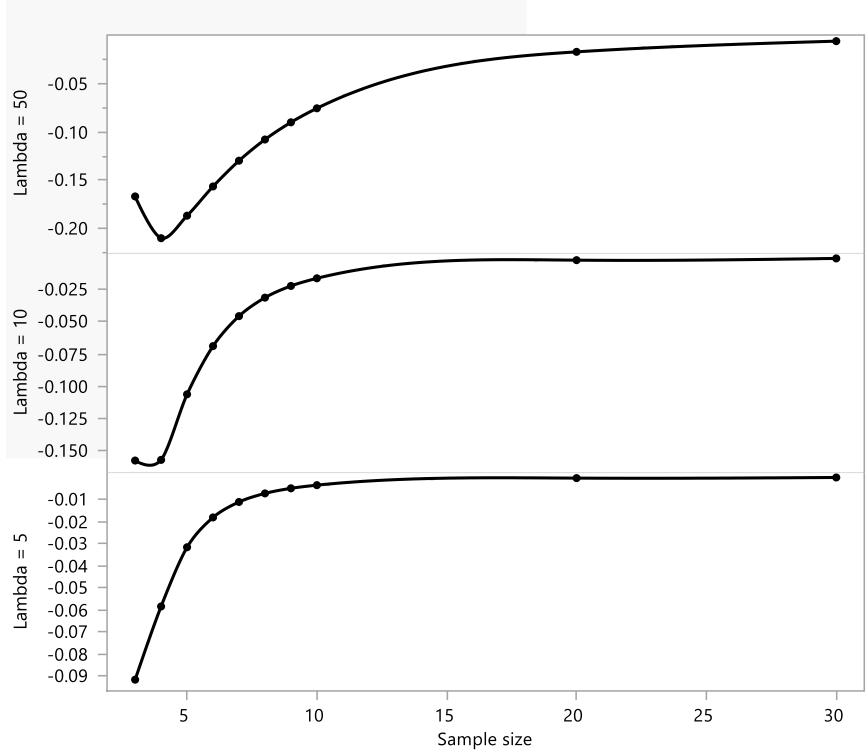

**Figure 11.** The derivative of $Var(\hat{\beta}_1)$ changing as a function of sample size $n$ (intercept and slope model) for $\lambda = 5, 10,$ and $50$.

*5.3 Covariance*

For the covariance, both the calculations and conclusions are relatively straight forward given the previous findings. From (3), $Cov(\hat{\beta}_0, \hat{\beta}_1) =$ $-\frac{6(1-\rho^2)(n-1)}{n^2(1+\rho^2-2\rho)+n(1+4\rho-5\rho^2)+6\rho(1+\rho)}$, which is equal to $-\frac{Var(\hat{\beta}_1)}{2}$. Consequently, the covariance has comparable conclusions to that which we reached when investigating the properties of $Var(\hat{\beta}_1)$; they just differ in sign. Table 10 highlights the $Cov(\hat{\beta}_0, \hat{\beta}_1)$ as both $n$ and $\lambda$ vary and Figure 12 shows this graphically.

**Table 10.** $Cov(\hat{\beta}_0, \hat{\beta}_1)$ with respect to $n$ and $\lambda$.

|  | | | Lambda ($\lambda$) | | | |
|---|---|---|---|---|---|---|
| Sample size ($n$) | 0.05 | 0.1 | 1 | 5 | 10 | 50 |
| 3 | -0.048771 | -0.095163 | -0.632121 | -0.993262 | -0.999955 | -1.000000 |
| 4 | -0.048771 | -0.095163 | -0.631856 | -0.954642 | -0.91506 | -0.900000 |
| 5 | -0.048771 | -0.095163 | -0.631742 | -0.932825 | -0.849233 | -0.800002 |
| 6 | -0.048771 | -0.095163 | -0.631686 | -0.920735 | -0.806075 | -0.714318 |
| 7 | -0.048771 | -0.095163 | -0.631654 | -0.913565 | -0.777936 | -0.643034 |
| 8 | -0.048771 | -0.095163 | -0.631634 | -0.909022 | -0.759051 | -0.583910 |
| 9 | -0.048771 | -0.095163 | -0.631622 | -0.905982 | -0.745931 | -0.534707 |
| 10 | -0.048771 | -0.095163 | -0.631613 | -0.903855 | -0.736515 | -0.493569 |
| 20 | -0.048771 | -0.095163 | -0.631587 | -0.897437 | -0.706886 | -0.306345 |
| 30 | -0.048771 | -0.095163 | -0.631582 | -0.896348 | -0.701665 | -0.256996 |
| 40 | -0.048771 | -0.095163 | -0.631581 | -0.895979 | -0.699888 | -0.238286 |
| 50 | -0.048771 | -0.095163 | -0.631580 | -0.895812 | -0.699079 | -0.229395 |

As one would expect, Table 10 suggests both an upper and lower bound for the covariance. Holding $n$ steady, $Cov(\hat{\beta}_0, \hat{\beta}_1) \to 0$ as $\lambda \to 0$ since $\rho \to 1$. However, the upper bound is a function of $n$, specifically $\frac{-6(n-1)}{n^2+n}$, as $\lambda \to \infty$ since $\rho \to 0$. To investigate the asymptotic value of $Cov(\hat{\beta}_0, \hat{\beta}_1)$, we take the limit as $n \to \infty$.

**Theorem 4:** $\lim_{n \to \infty} Cov(\hat{\beta}_0, \hat{\beta}_1) = -\frac{12\lambda}{\lambda^2+6\lambda+12} = \frac{-2}{\frac{\lambda}{6}+1+\frac{2}{\lambda}}$

**Proof:** $\lim_{n \to \infty} Cov(\hat{\beta}_0, \hat{\beta}_1) = \lim_{n \to \infty} -\frac{Var(\hat{\beta}_1)}{2} = \left(\frac{-1}{2}\right)\left(\frac{4}{\frac{2}{\lambda}+1+\frac{\lambda}{6}}\right) = \frac{-2}{\frac{\lambda}{6}+1+\frac{2}{\lambda}}.$  Q.E.D.

As anticipated, the $Cov(\hat{\beta}_0, \hat{\beta}_1)$ converges to a non-zero value that is a function of $\lambda$. In comparison, under the customary IID assumptions of the increasing domain structure, $Cov(\hat{\beta}_0, \hat{\beta}_1) = \frac{-\bar{X}\sigma^2}{\sum_{i=1}^{n}(X_i-\bar{X})^2} = -\bar{X}Var(\hat{\beta}_1)$, which converges to 0 as $n \to \infty$. With the limiting covariance equation under the IA domain, it is noted that 0 is indeed both the upper bound as $\lambda \to 0$ and $\lambda \to \infty$. The minimum value for this curve, -.9282, is attained at $\lambda = \sqrt{12}$. Figure 13 illustrates this pattern. This figure also indicates the maximum impactful $\lambda$ value. Here, the criterion is adjusted in order to find the $\lambda$ value as the point at which the asymptotic variance increases by <.0001 (in absolute value) for a 0.1 increase in $\lambda$. to only account for values of $\lambda$ that are greater than $\sqrt{12}$. Under this model, the maximum $\lambda$ value needed is 103.7.

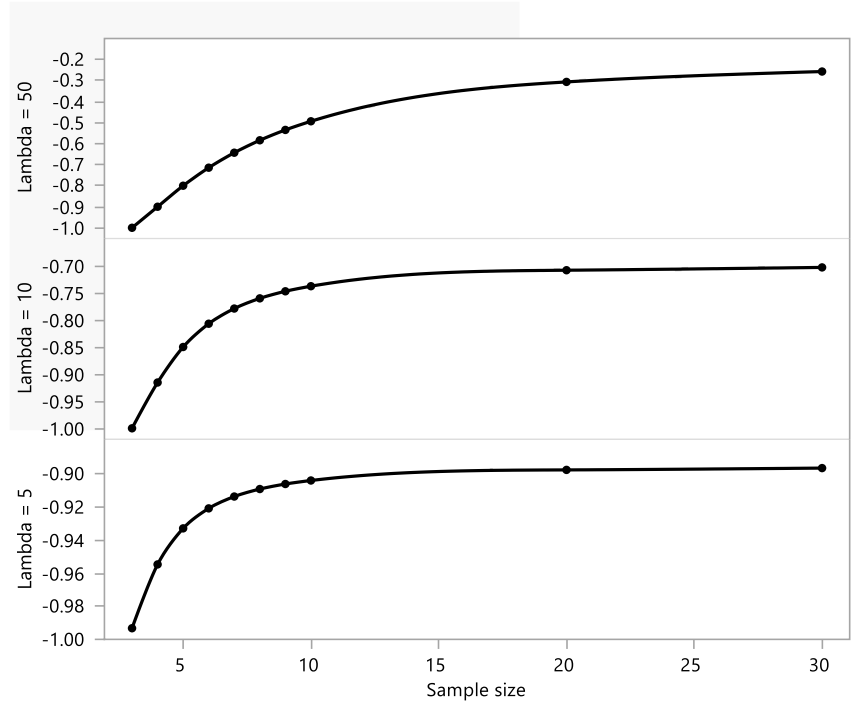

**Figure 12.** $Cov(\hat{\beta}_0, \hat{\beta}_1)$ versus sample size, *n*, for the intercept and slope model for $\lambda = 5, 10,$ and 50.

To ascertain the rate at which $Cov(\hat{\beta}_0, \hat{\beta}_1) \to \frac{-2}{\frac{\lambda}{6}+1+\frac{2}{\lambda}}$, we take the derivative of $Cov(\hat{\beta}_0, \hat{\beta}_1)$ with respect to *n*, the values of which are equal to those derived for $Var(\hat{\beta}_1)$ but simply divided by -2. Table 11 highlights how quickly this occurs, while Figure 14 shows this notationally for select values of $\lambda$.

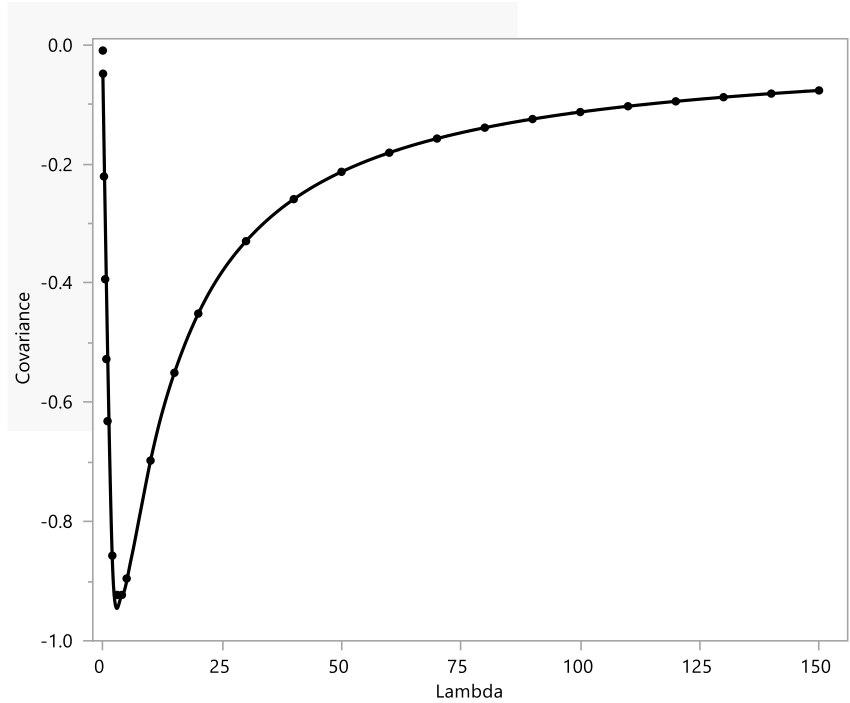

**Figure 13.** Asymptotic $Cov(\hat{\beta}_0, \hat{\beta}_1)$, for the intercept and slope model, as a function of $\lambda$.

**Table 11.** Derivative evaluations of $Cov(\hat{\beta}_0, \hat{\beta}_1)$, intercept and slope model, as $n$ and $\lambda$ vary.

|  |  |  | Lambda ($\lambda$) |  |  |  |
|---|---|---|---|---|---|---|
| Sample size ($n$) | 0.05 | 0.1 | 1 | 5 | 10 | 50 |
| 3 | 3.10E-10 | 9.43E-09 | 0.000395 | 0.045820 | 0.079069 | 0.083333 |
| 4 | 1.26E-10 | 3.83E-09 | 0.000165 | 0.029244 | 0.078812 | 0.105000 |
| 5 | 5.81E-11 | 1.77E-09 | 7.70E-05 | 0.015848 | 0.053247 | 0.093333 |
| 6 | 3.09E-11 | 9.42E-10 | 4.12E-05 | 0.009098 | 0.034459 | 0.078231 |
| 7 | 1.83E-11 | 5.56E-10 | 2.44E-05 | 0.005602 | 0.022771 | 0.065051 |
| 8 | 1.16E-11 | 3.55E-10 | 1.56E-05 | 0.003662 | 0.015565 | 0.054397 |
| 9 | 7.86E-12 | 2.40E-10 | 1.05E-05 | 0.002513 | 0.011007 | 0.045922 |
| 10 | 5.55E-12 | 1.69E-10 | 7.43E-06 | 0.001794 | 0.008026 | 0.039161 |
| 20 | 5.99E-13 | 1.83E-11 | 8.04E-07 | 0.000200 | 0.000954 | 0.011631 |
| 30 | 1.69E-13 | 5.15E-12 | 2.27E-07 | 5.67E-05 | 0.000273 | 0.004779 |
| 40 | 6.95E-14 | 2.12E-12 | 9.33E-08 | 2.34E-05 | 0.000113 | 0.002325 |
| 50 | 3.51E-14 | 1.07E-12 | 4.71E-08 | 1.18E-05 | 5.72E-05 | 0.001278 |

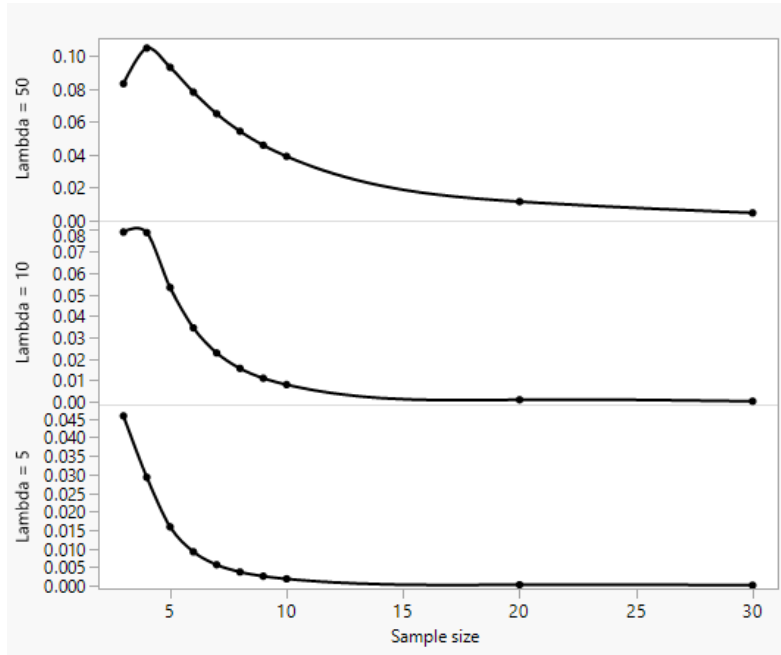

**Figure 14.** The derivative of $Cov(\hat{\beta}_0, \hat{\beta}_1)$ changing as a function of sample size $n$ (intercept and slope model) for $\lambda = 5, 10,$ and $50$.

As expected, the same two characteristics seen before holding true for the covariance. The covariance is converging relatively quickly to $\frac{-2}{\frac{\lambda}{6}+1+\frac{2}{\lambda}}$ for small to moderate values of $\lambda$ and there appears to be a point of diminishing return as sample size increases for all values of $\lambda$.

## 6. CONCLUSION

A fixed time domain differs from the traditional statistical approach of an infinite observational domain. Consequently, an appropriate stochastic error should be adopted to account for the increased correlation between observations as sample size increases. Through the adoption of the Ornstein-Uhlenbeck process, or a spatial AR(1), we properly accounted for this serial correlation.

In utilizing (1) through adopting a simple linear regression, we demonstrate that although (2) results in estimators that are both BLUE and UMVUE for $\boldsymbol{\beta}$, they exhibit variances or covariances that do not converge to zero as sample size tends to infinity. Consequently, we prove the estimators in this article are inconsistent and extend the IA results of Morris and Ebey (1984), Lahiri (1996), White (2001), and Mills (2010) for simple linear regression models. Moreover, our results extend to any smooth function using piecewise linear functions (Sontag 1981). That is, any function that can be approximated by a combination of linear functions will have a set of parameters that are unbiased but inconsistent as sample size tends to infinity in an IA domain.

Among each of the estimator results, a shared property is the importance of the tuning parameter, $\lambda$. As $\lambda \to \infty$ (and assuming $n \to \infty$) all estimator variances tend to zero, mirroring the results of consistent estimators under an increasing time domain customarily assumed in statistical asymptotic theory. Other patterns also emerge from the results. For a given $\lambda$, estimator variance decreases as sample size increases. Consequently, the derivatives of the parameter variances approach zero at a much faster rate for smaller $\lambda$. Lastly for a given $n$, estimator variances generally decrease as $\lambda$ increase. Additionally, points of diminishing returns on an increasing $\lambda$ for the asymptotic variance of the parameter estimators are indicated for each model. The largest $\lambda$ value noted is 148.9 that indicates utilizing a $\lambda$ above this value is unnecessary for the given models.

Through the derivation of the exact variance of each parameter estimator (intercept or slope), we also display that the variance is a function of both $n$ and the tuning parameter, $\lambda$. With respect to $n$, each estimator variance appears to have a point of diminishing return. Although none of the variances approach zero, they display a pattern of greater diminishing return regarding decreasing estimator variance as *n* increases. This is invaluable to a practitioner as this indicates perhaps an optimal sample size to cease data collection. This in turn reduces time and data collection cost for little information is gained in sampling beyond a certain sample size. This article does not explicitly explore what this value is, but future research can investigate this concept to determine an optimal sample size.